\documentclass[a4paper,twoside]{article}      
\usepackage{amsmath,amssymb,amsfonts} 
\usepackage{graphics}                 
\usepackage{color}                    
\usepackage{hyperref}                 
\usepackage{floatflt,epsfig}
\usepackage{amsmath}
\usepackage{empheq}
\definecolor{myblue}{rgb}{.8, .8, 1}
\definecolor{myblue2}{rgb}{.9, .9, 0.9}
\definecolor{myblue3}{rgb}{.5, .9, 0.9}

\newlength\mytemplen
\newsavebox\mytempbox

\makeatletter
\newcommand\mybluebox{%
    \@ifnextchar[
       {\@mybluebox}%
       {\@mybluebox[0pt]}}

\def\@mybluebox[#1]{%
    \@ifnextchar[
       {\@@mybluebox[#1]}%
       {\@@mybluebox[#1][0pt]}}

\def\@@mybluebox[#1][#2]#3{
    \sbox\mytempbox{#3}%
    \mytemplen\ht\mytempbox
    \advance\mytemplen #1\relax
    \ht\mytempbox\mytemplen
    \mytemplen\dp\mytempbox
    \advance\mytemplen #2\relax
    \dp\mytempbox\mytemplen
    \colorbox{myblue2}{\hspace{1em}\usebox{\mytempbox}\hspace{1em}}}

\makeatother

\def\beq{\begin{equation}}
\def\eeq{\end{equation}}
\def\beqn{\begin{eqnarray}}
\def\eeqn{\end{eqnarray}}
\newcommand{\be}{\begin{eqnarray}}
\newcommand{\ee}{\end{eqnarray}}

\def\R{\mathbb{ R}}
\def\S{\mathbb{ S}}

\date{\small \it April 12, 2013}
\title{{\bf \color{black}
Finite Quantum Gravity}
}

\author{Leonardo Modesto \\
{\small Department of Physics \& Center for Field Theory and Particle Physics,} \\
{\small Fudan University, 200433 Shanghai, China}
 }

\begin{document}
\maketitle
\begin{abstract}
{ 
We hereby present a class of multidimensional higher derivative theories of gravity 
that realizes an ultraviolet completion of Einstein general relativity. This class is
marked by a ``non-polynomal" entire function (form factor), which averts
extra degrees of freedom (including ghosts) and improves the high energy
behavior of the loop amplitudes. By power counting arguments, it is
proved that the theory is super-renormalizable in any dimension, i.e.
only one-loop divergences survive. Furthermore, in odd dimensions there
are no counter terms for pure gravity and the theory turns out to be
``finite." Finally, considering the infinite tower of massive
states coming from dimensional reduction, quantum gravity is finite in
even dimension as well. 
%
%
%
%
}
\end{abstract}

{\color{blue}{\sc keywords: Quantum Gravity, Extra Dimensions.}} 

\tableofcontents

\section{Introduction}
The greatest physicists of the of the 20th century were able to find a consistent quantum field theory 
for all fundamental interactions except for gravity. 
Starting from quantum electrodynamics and 
quantum non-abelian gauge theories up to the standard model of particle physics,
two guiding principles seem to dominate the research in high-energy physics: 
{\em renormalization and perturbative theory}. 
However, gravity seems to elude so far these patterns and many authors suggest ingenious solutions to one of the biggest puzzles of our days, but none is completely satisfactory.
But first and foremost, are we really sure about the quantum nature of gravity at very short distance?
There are  many reasons to believe that 
gravity has to be quantum, some of which are: the quantum nature of matter in the right-hand side of the 
Einstein equations, the singularities appearing in classical solutions of general relativity, etc. 
However, the major obstacle when we try to construct a consistent theory of quantum gravity is that
Einstein's dynamics is ``non-renormalizable" by conventional criteria and hence it is not 
capable to tame in any way the ambiguous predictions coming out at quantum level.
It is common belief that general relativity and quantum mechanics are not compatible, but there is nothing inconsistent between them. 
Just like Fermi's theory of weak interactions, quantum Einstein's gravity is solid and calculable. 
As mentioned above, it is only non-renormalizable and therefore non-perturbative for energies $E \gtrsim M_{\rm P}$ (Planck mass), but it is not inconsistent. At short distances, higher order 
operators in the Lagrangian become decisive. Therefore, if we want to use a diffeomorphism invariant theory 
for a massless spin two particle at short distance, we need an ultraviolet completion of Einstein's gravity. 
None of the solutions provides so far has been completely successful. 

The aim of this work is to extend Einstein's general relativity 
to make gravity compatible with the above guiding principles (renormalization and perturbative theory) 
in the ``quantum field theory framework". 
In this paper we will work in a multidimensional spacetime:
I
improving the ultra-violet behavior of a theory by increasing 
the number of spacetime dimensions may sound paradoxical, but not if we are looking for finiteness in addition to
power counting renormalizability.  As a matter of fact, the chances of finiteness seem to be 
better in odd dimension, prompting us to look for a theory of quantum gravity in extra dimensions.

In some recent papers \cite{modesto,BM,Bis2, M2, M3, M4}, 
a different action principle for gravity has been introduced 
to make up for the shortcomings of the quantization of the Einstein-Hilbert action.
The theory fulfills 
a synthesis of minimal requirements: 
{\color{black}
\begin{enumerate}
\renewcommand{\theenumi}{(\roman{enumi})}
\item 
at classical level 
we assume as a guiding principle the regularity of the space-time at every scale; in other words classical solutions must be singularity-free \cite{ModestoMoffatNico,Bis3};  
\item Einstein-Hilbert action should be a good approximation of the theory at a much smaller energy scale than the Planck mass;  
\item the spacetime dimension has to decrease with the energy in order to have 
a complete quantum theory in the ultraviolet regime; 
\item the theory has to be perturbatively super-renormalizable 
or finite at quantum level; 
\item the theory has to be unitary, with no other 
pole in the propagator in addition to the graviton; 
\item spacetime is a single continuum of space and time and in particular the 
Lorentz invariance is not violated. 
\end{enumerate}
}
This approach to quantum gravity is partly inspired by Cornish and Moffat's papers \cite{Moffat,corni1,corni2,Moffat2,corni3,Moffat4}
and mainly inspired by Tomboulis' studies on super-renormalizable gauge theories \cite{Tombo} and quantum gravity \cite{Tombo70, Tombo80, TomboAnto}.
Research records show that Krasnikov proposed a similar theory in 1989 \cite{Krasnikov},
following Efimov's studies \cite{efimov,E2,E3,E4,E5}.

%
{\color{cyan}
{\bf Definitions}. 
The definitions used in this paper are: the metric tensor $g_{M N}$ has 
signature $(+ - \dots -)\,$; the curvature tensor  
$R^{M}_{N P Q } = - \partial_{Q} \Gamma^{M}_{N P} + \dots$, 
the Ricci tensor $R_{M N} = R^{P}_{M N P}$, and the curvature scalar $R = g^{M N} R_{M N}$. 
%
}

\section{Multidimensional 
quantum gravity}

The class of theories we are going to introduce 
is a 
``non-polynomial"
extension of the renormalizable quadratic Stelle theory \cite{Stelle}
and it has the following general structure  \cite{modesto, BM},  
\be
&& \hspace{-0.65cm} 
 \mathcal{L} =  
2 \kappa^{-2} R+ R_{M N}  \, \gamma_2(\Box)  R^{M N} \! + R \,\gamma_0( \Box)  R 
+ R_{MNPQ} \gamma_4(\Box) R^{MNPQ} 
+ 
O(R^3) +\dots + R^{D/2} 
 \, ,  \nonumber 
\ee
where 
the three ``form factors" $\gamma_{0,2,4}(\Box)$ are ``entire functions" of the covariant D'Alembertian operator, $\Box := \Box/\Lambda^2$, $\Lambda$ is an invariant mass scale
and $\kappa^2 = 32 \pi G_N$ in $D=4$.
The non-polynomiality only involves positive powers of the D'Alembertian operator since the two form 
factors are entire functions. 
The theory is not unique, but all the freedom present in the action can be read in 
the three functions $\gamma_{0,2,4}$ \cite{Tombo, Krasnikov, efimov}. 


In 1977 Stelle introduced and studied a four dimensional higher derivative quantum gravity \cite{Stelle, Shapirobook}, whose generalization in a $D$-dimensional spacetime reads 
\be
S = 
\int {\rm d}^D x \sqrt{-g} \Big[ \frac{2}{ \kappa^{2}} R + \frac{\alpha}{2} R^2 + \frac{\beta}{2} R_{M N} R^{M N}\Big].
\label{OldStelle}
\ee
If we calculate the upper bound to the superficial degree of freedom for this theory, 
we find 
\be
\delta= D L - 4 I + 4 V  
= D - (D-4)(V-I) 
= D - (D-4)(V-I) = D + (D-4)(L-1) \, ,
\label{deltaD}
\ee
where $L$ is the number of loops, 
$V$ is the number of vertices and $I$ the number of internal lines of the graph.
We have substituted in $\delta$ the topological relation 
$L = 1+ I -V$.
In $D=4$ we get $\delta =4$
and the theory is then 
renormalizable, since all the divergences can be absorbed in the operators already present in 
the Lagrangian (\ref{OldStelle}). On the contrary, for $D>4$ the theory (\ref{OldStelle}) is non renormalizable. 
Unfortunately, however, the propagator contains a physical 
ghost (state of negative norm) that represents  
a violation of unitarity. Probability, as described by the scattering $S$-matrix, is no longer preserved.
Similarly, the classical theory is destabilized, 
since the dynamics can drive the system to become arbitrarily excited,
and the Hamiltonian constraint is unbounded from below. 
On this basis, we can generalize the Stelle theory to a theory renormalizable in any dimension, so that 
the Lagrangian with at most $X$ derivatives of the metric is 
\be 
&& \hspace{-0.7cm} 
{\mathcal L}_{D-{\rm Ren} } = a_1 R + a_2 R^2 + b_2 R_{M N}^2 + 
\label{localDren} 
 \dots + a_X R^{\frac{X}{2}} + b_X R_{M N}^{\frac{X}{2}} +
c_X R_{M N PQ}^{\frac{X}{2}} + d_X R \, \Box^{\frac{X}{2} - 2} R \dots  \, . 
\ee
The middle dots stand for a finite number of extra 
terms with fewer derivatives of the metric tensor, and the dots on the right indicate 
a finite number of operators with the same 
number of derivatives but higher powers of the curvature $O(R^2 \Box^{X/2-3} R)$.

In this theory, the power counting tells us that the maximal superficial degree of divergence 
of a Feynmann graph is 
\be
\delta = D - (D - X)(V  -  I) = D + (D - X)(L  -  1).
\label{DDX}
\ee
For $X = D$, the maximal divergence is $\delta =D$: this assumption makes the theory renormalizable and  
all the infinities can be absorbed in the operators already present in the action (\ref{localDren}).

In 1996, Asoreya, L$\grave{{\rm o}}$pez and Shapiro \cite{shapiro} contributed to evolve these two stages of 
the theory towards a local ``super-renormalizable gravity".  The theory including all general covariant 
operators with up to $2\mathrm{N}+4$ derivatives, reads 
\be 
\hspace{0.2cm} 
S = \!\! \int \! d^D x \sqrt{|g|} \Big[2\, \kappa^{-2} \, R + \bar{\lambda}
+ \sum_{n=0}^{\mathrm{N}} \! \Big( 
a_n \, R \, \Box^n  R  + 
b_n \, R_{M N} \, \Box^n \, R^{M N} 
\Big) 
+ c_1^{(1)} R^3+ \dots + c_1^{(\mathrm{N})}R^{\mathrm{N}+2} \Big]. \hspace{-0.0cm}
 \label{actionshapi}
\ee 
In the action above, we distinguished operators at most quadratic in the curvature, from operators
$O(R^3)$. 
To make the the spacetime dimension $D$ explicit, 
we introduce another integer $\mathrm{M}$ 
in the curvature expansion: $\mathrm{N} = \mathrm{M} + D_{\rm even}/2$ or
$\mathrm{N} = \mathrm{M} + (D_{\rm odd}+1)/2$. 
The maximal superficial degree of divergence of a Feynmann amplitude with $L$-loops reads
\be
&& \hspace{-0.7cm} 
\delta \leqslant 
    L D_{\rm even} - (2 \mathrm{N} +4) (I - V) 
=  L D_{\rm even}  - (2 \mathrm{M} + D_{\rm even}  + 4) (L-1) 
=  D_{\rm even}  - (2 \mathrm{M} +4) (L-1) , \nonumber \\
&&  \hspace{-0.7cm} 
\delta \leqslant 
    L D_{\rm odd} - (2 \mathrm{N} +4) (I - V) 
=  L D_{\rm odd}  - (2 \mathrm{M} + D_{\rm odd}  +5) (L-1) 
=  D_{\rm odd}  - (2 \mathrm{M} +5) (L-1).
\ee
Clearly, only one loop divergences survives for $\mathrm{M}> (D_{\rm even} -4)/2$, $\mathrm{M}> (D_{\rm odd} - 5)/2$  or ${\mathrm N}>D-2$. 
Therefore the theory is super-renormalizable but still non-unitary.

\subsection{Multidimensional super-renormalizable gravity}
Given the previous work on local super-renormalizable gravity, we are now going 
to address the unitarity together with the renormalizability with a different action. 
In a $D$-dimensional spacetime,
the general action of ``derivative order $N$" can be found combining curvature tensors with their covariant derivatives. In a nutshell, the action reads 
as follows \cite{shapiro},
\be
\!\!\!\! S = \sum_{n=0}^{\mathrm{ N }+2} \alpha_{2 n} \Lambda^{D - 2 n} \int d^D x \sqrt{ | g |} \, \mathcal{O}_{2 n}  (\partial_{Q } g_{M N}) + S_{\rm NP} \, ,
\label{generalAction}
\ee
where $\Lambda$ is an invariant mass scale in our fundamental theory, 
$\mathcal{O}_{2n} (\partial_{Q} g_{M N})$ denotes the general covariant scalar term containing 
``$2 n$" derivatives of the metric $g_{\mu \nu}$, while $S_{\rm NP}$ is a 
non-polynomial action term that 
we are going to set 
later \cite{Tombo}. The maximal number of derivatives in the local part of the action is $2 \mathrm{N} +4$.
We can then classify the local terms in the following way,
\be
&& 
\mathcal{O}_0 = \{ \lambda \} \, , \,\, \nonumber \\
&&
  \mathcal{O}_2 = \{ R \} \, , \,\, \nonumber \\
  && 
  \mathcal{O}_4 = \{ R^2, \, R_{M N} R^{M N} , \,\, 
R_{M N P Q} R^{M N P Q} \} \, , \,\, \nonumber \\
&&
\mathcal{O}_6 = \{R^3_{\dots}, \, \nabla R_{\dots} \nabla R_{\dots}, \dots  \} \, , \label{operators} \\
&&  
\mathcal{O}_8 = \{ R^4_{\dots}, \,  R_{\dots} \nabla R_{\dots} \nabla R_{\dots}, \,  \nabla^2 R_{\dots} \nabla^2 R_{\dots} \, , \dots \} \, , \,\, \nonumber \\
&& \dots \nonumber \\
&&
\mathcal{O}_N = \{ R^{N+2}_{\dots}, \, R^{N-1}_{\dots} \nabla R_{\dots} \nabla R_{\dots}, \,  
R_{\dots} \Box^N R_{\dots}, \dots \} . \nonumber 
\ee
In the local theory (\ref{localDren}), renormalizability 
requires $X = D$, so that the relation between
the spacetime dimension and the derivative order is $2\mathrm{N} +4 = D$.
To avoid fractional powers of the D'Alembertian operator, we take
$2 \mathrm{N} + 4 = D_{\rm odd} +1$ in odd dimensions and $2 \mathrm{N} + 4 = D_{\rm even}$
in even dimensions.
Given the general structure (\ref{generalAction}), for $\mathrm{N} \geqslant 0$ and $n \geqslant 2$ 
contributions to the propagator come only from the following operators, 
\be
R_{M N} \Box^{n-2} R^{M N} \, , \,\,\, 
R  \, \Box^{n-2} R \, , \,\,\, 
R_{M N P Q } \Box^{n-2} R^{M N P Q}.
\label{terms}
\ee
However, using the Bianchi and Ricci identities, we can scale the three terms listed above
down to two, 
\be
R_{M N P Q  } \Box^{n-2} R^{M N P Q }    
= 4  R_{M N} \Box^{n-2} R^{M N} 
- R \,  \Box^{n-2} R + O(R^3) + \nabla_{P} \Omega^{P} , 
\label{property0}\ee
where 
$\nabla_{P} \Omega^{ P }$ 
is a total divergence term.
Applying 
(\ref{property0})
to (\ref{terms}), for $n \geqslant 2$ we discard the third term and we keep the first two along with higher curvature terms.

We now have to define the 
non-polynomial action term in (\ref{generalAction}). 
As we are going to show, both super-renormalizability and unitarity require  
two non-polynomial operators,
\be 
 R_{M N} \,  h_2( - \Box_{\Lambda}) R^{M N} \, , \,\,\, \, 
  R   \, h_0( - \Box_{\Lambda}) \, R \, .
\ee 
The full action, focusing mainly on the non-local terms and on the quadratic part in the curvature, reads
\be 
&& \hspace{-1.5cm} 
S = \int d^D x \sqrt{|g|} \Big[2\, \kappa^{-2} \, R + \bar{\lambda}
+ \sum_{n=0}^{\mathrm{N}} \Big( 
a_n \, R \, (-\Box_{\Lambda})^n \, R  + 
b_n \, R_{M N} \, (-\Box_{\Lambda})^n \, R^{M N} 
\Big) 
\nonumber \\
&& 
+ R  \, h_0( - \Box_{\Lambda}) \, R +
R_{M N} \, h_2( - \Box_{\Lambda}) \, R^{M N}
+ \, \underbrace{ O(R^3) \dots\dots\dots + R^{\mathrm{N}+2}}_{\mbox{Finite number of terms}} \Big] 
 \, . \label{action}
\ee 
The last line is a collection of local terms that are renormalized at quantum level.
In the action, 
 the couplings and the non-polynomial functions have the following dimensions: 
 \be
 [a_n] =[b_n]= M^{D-4} \, , \,\,\,\, [\kappa^2] = M^{2-D} \, , \,\,\,\,
 [\bar{\lambda}]=M^D \, ,  \,\,\,\, [h_2] = [h_0] = M^{D-4}.
 \ee
 
 \subsection{Graviton propagator} \label{gravitonpropagator}
At this point, we are ready to expand the Lagrangian at the second order in the graviton fluctuation.
Splitting the spacetime metric in the flat Minkowski background and the fluctuation $h_{M N}$ 
defined by $g_{M N} =  \eta_{M N} + \kappa \, h_{M N}$,
we get \cite{HigherDG}
 %
\be
&& \hspace{-1cm} \mathcal{L}_{\rm lin} = 
- \frac{1}{2} [ h^{M N} \Box h_{M N} + A_{N}^2 + (A_{N} - \phi_{, N})^2 ] 
+ \frac{1}{4} \Big[ \frac{\kappa^2}{2} \Box h_{M N}  \beta( \Box) \Box h^{M N} 
- \frac{\kappa^2}{2} A^{M}_{, M}  \beta( \Box) A^{N}_{, N} 
 \nonumber \\
&&  \hspace{-1cm} - \frac{\kappa^2}{2} F^{M N}  \beta( \Box) F_{M N} 
+ \frac{\kappa^2}{2} (A^{P}_{, P} - \Box \phi) \beta( \Box) (A^{Q}_{, Q} - \Box \phi)
+ 2 \kappa^2 \left(A^{P}_{, P}  - \Box \phi \right) \alpha( \Box) (A^{Q}_{, Q} - \Box \phi ) \Big]  
\,  ,
\label{quadratic2} 
\ee
where $A^{M} = h^{M N}_{\,\,\,\, , N}$, $\phi = h$ (the trace of $h_{M N}$), 
$F_{M N} = A_{M , N} - A_{N, M}$ and the functionals of the D'Alembertian operator 
$\beta (\Box), \alpha(\Box)$ are defined by 
\be
\alpha(\Box)  :=  2  \sum_{n = 0}^{\mathrm{N} } a_n ( - \Box_{\Lambda})^n + 2 h_0(- \Box_{\Lambda}) , 
\,\,\,\,\,\,
 \beta(\Box)  :=  2  \sum_{n = 0}^{\mathrm{N}} b_n ( - \Box_{\Lambda})^n + 2 h_2(- \Box_{\Lambda})   .
 \label{alphabeta}
\ee
The d'Alembertian operator in $\mathcal{L}_{\rm lin}$ and (\ref{alphabeta}) must be conceived on 
the flat spacetime. 
The linearized Lagrangian (\ref{quadratic2}) is invariant under infinitesimal coordinate transformations 
$x^{M} \rightarrow x^{M} + \kappa \, \xi^{M}(x)$, where $\xi^{M}(x)$ is an infinitesimal vector field 
of dimensions $[\xi(x)] = M^{(D-4)/2}$. Under this transformation, the graviton field turns into 
$h_{M N} \rightarrow h_{M N} - \xi(x)_{M, N} - \xi(x)_{N, M}.$
The presence of the local gauge simmetry 
calls for the addition of a gauge-fixing term
to the linearized 
Lagrangian (\ref{quadratic2}). Hence, we choose the following fairly general gauge-fixing operator
\be
 && \hspace{0cm} \mathcal{L}_{\rm GF}  =  \lambda_1 (A_{N} - \lambda \phi_{,N}) \omega_1(-\Box_{\Lambda}) (A^{N} - \lambda \phi^{,N})
\nonumber \\
&& 
+ \frac{\lambda_2 \, \kappa^2}{8} (A^{M}_{,M} - \lambda \Box \phi) \beta(\Box) \omega_2(-\Box_{\Lambda}) (A^{N}_{,N} - \lambda \Box \phi)
+ \frac{\lambda_3 \, \kappa^2}{8}    
F_{M N} \,
\beta(\Box) \omega_3(-\Box_{\Lambda}) \, 
 F^{M N}  \, ,
 \label{GF2}
\ee
where $\omega_i( - \Box_{\Lambda})$ are three weight functionals \cite{Stelle}. 
In the harmonic gauge 
$\lambda=\lambda_2 = \lambda_3 = 0$ and 
$\lambda_1 = 1/\xi$.
The linearized gauge-fixed Lagrangian reads 
\be
\mathcal{L}_{\rm lin} + \mathcal{L}_{\rm GF} = 
\frac{1}{2} h^{M N} \mathcal{O}_{M N, P Q} \, h^{P Q},
\label{O}
\ee
where the operator 
$\mathcal{O}$ is made of two terms, one coming from the linearized 
Lagrangian 
(\ref{quadratic2}) and the other from the gauge-fixing term (\ref{GF2}).
Inverting the operator $\mathcal{O}$ \cite{HigherDG}, we find the 
two-point function in the harmonic gauge ($\partial^{M} h_{M N} = 0$),
\be
\mathcal{O}^{-1}(k) = \frac{\xi (2P^{(1)} + \bar{P}^{(0)} ) }{2 k^2 \, \omega_1( k^2/\Lambda^2)} 
+ \frac{P^{(2)}}{k^2 \Big(1 + \frac{k^2 \kappa^2 \beta(k^2)}{4} \Big)} 
- \frac{P^{(0)}}{2 k^2 \Big( \frac{D-2}{2} - k^2 \frac{D \beta(k^2) \kappa^2/4 + (D-1) \alpha(k^2) \kappa^2}{2} \Big) } \,  . \label{propagator}
\ee
The tensorial 
indexes for the operator $\mathcal{O}^{-1}$ and the projectors $P^{(0)},P^{(2)},P^{(1)},\bar{P}^{(0)}$ have been omitted and the functions $\alpha(k^2)$ and $\beta(k^2)$ are achieved by replacing $-\Box \rightarrow k^2$ in the definitions (\ref{alphabeta}). The projectors are defined by 
\cite{HigherDG, VN}\label{proje2}
{\small
\be
\hspace{0.5cm}
\boxed{
\begin{array}{rcl}
 && 
 P^{(2)}_{M N P Q }(k) = \frac{1}{2} ( \theta_{M P} \theta_{N Q } +
 \theta_{M Q} \theta_{N P} ) - \frac{1}{D-1} \theta_{M N} \theta_{P Q} \,  ,
 \nonumber \\
&&    P^{(1)}_{M N P Q }(k) = \frac{1}{2} \left( \theta_{M P} \omega_{N Q} +
 \theta_{M Q} \omega_{N P}  + 
 \theta_{N P} \omega_{M Q}  +
  \theta_{N Q} \omega_{M P}  \right) ,
\\
 && 
 P^{(0)} _{M N P Q } (k) = \frac{1}{D-1}  \theta_{M N} \theta_{P Q}  , \,\,\,\, \,\,
\bar{P}^{(0)} _{M N P Q } (k) =  \omega_{M N} \omega_{P Q} \, , \,\,\,\,\,  \nonumber \\
&& 
\bar{\bar{P}}^{(0 )} _{M N P Q }  = \theta_{M N} \omega_{P Q }
+ \omega_{M N} \theta_{P Q} \, , \,\,\,\,\, 
\theta_{M N } = \eta_{M N} - \frac{k_{M } k_{N }}{k^2}  \, , \,\,\,\,\,\,
 \omega_{M N} = \frac{k_{M} k_{N}}{k^2} \, . \nonumber 
 \end{array}
 }
\ee
}
%

By looking at the last two gauge invariant terms in (\ref{propagator}), we deem convenient to introduce the following definitions, 
\be
\bar{h}_2(z) \!\!\!\! & = & \!\!\!\! 1 + \frac{\kappa^2 \Lambda^2}{2}  z \sum_{n=0}^{\mathrm{N}} b_n z^n + \frac{\kappa^2 \Lambda^2}{2}
z \, h_2(z) \, , \label{barh2h0}\\
\left( \frac{D-2}{2} \right) \! \bar{h}_0(z) \!\!\!\! & = & \!\!\!\!  \frac{D-2}{2} - \frac{\kappa^2 \Lambda^2 D}{4}  z 
\left[\sum_{n=0}^{\mathrm{N}} b_n z^n + h_2(z) \right] 
- \kappa^2 \Lambda^2 (D - 1)  z \left[\sum_{n=0}^{\mathrm{N}}a_n z^n + h_0(z) \right] , \nonumber 
\ee
where again $z = - \Box_{\Lambda}$. 
Through the above definitions (\ref{barh2h0}),
the gauge invariant part of the propagator greatly simplifies to
%
%
\be
 \mathcal{O}^{-1}(k)
 = \frac{1}{k^2}
\left( \frac{P^{(2)}}{\bar{h}_2} 
- \frac{P^{(0)}}{(D-2) \bar{h}_0 } \right) + {\rm gauge} \,\, {\rm terms}.
\label{propgauge}
\ee


Once established that $h_2$ and $h_0$ are not polynomial functions, we demand 
the following general properties for the transcendental entire functions $h_i(z)$ ($i = 0,2$) and/or 
$\bar{h}_i(z)$ ($i = 0,2$)
\cite{Tombo}:
\begin{enumerate}
\renewcommand{\theenumi}{(\roman{enumi})}
\renewcommand{\labelenumi}{\theenumi}
\item $\bar{h}_i(z)$ ($i=0,2$) is real and positive on the real axis and it has no zeroes on the 
whole complex plane $|z| < + \infty$. This requirement implies that there are no 
gauge-invariant poles other than the transverse massless physical graviton pole.
\item $|h_i(z)|$ has the same asymptotic behavior along the real axis at $\pm \infty$.
\item There exists $\Theta>0$ such that 
\be
&& \lim_{|z|\rightarrow + \infty} |h_i(z)| \rightarrow | z |^{\gamma + \mathrm{N}}, \nonumber \\
&& \gamma\geqslant D/2 \,\,\,\, {\rm for} \,\,\,\, D = D_{\rm even}  
\,\,\,\, {\rm and} \,\,\, \, 
\gamma\geqslant (D-1)/2 \,\,\,\, {\rm for} \,\,\,\, D = D_{\rm odd} \, , 
\label{tombocond}
\ee 
for the argument of $z$ in the following conical regions  
\be
C = \Big\{ z \, | \,\, - \Theta < {\rm arg} z < + \Theta \, , \,\,  \pi - \Theta < {\rm arg} z < \pi + \Theta \Big\} , \,\,\,\, 
{\rm for } \,\,\, 0< \Theta < \pi/2. \nonumber 
\ee
This condition is necessary to achieve the super-renormalizability of the theory that we 
are going to show here below. The necessary 
asymptotic behavior is imposed not only on the real axis, (ii) but also on the conic regions that surround it.  
In an Euclidean spacetime, the condition (ii) is not strictly necessary if (iii) applies.
\end{enumerate}

\subsection{Power counting of loop diagrams and renormalization}
Let us then examine the ultraviolet behavior of the quantum theory.
According to the property (iii) in the high energy regime, the propagator in the momentum space goes as 
\be
\mathcal{O}^{-1}(k) \sim \frac{1}{k^{2 \gamma +2 \mathrm{N} +4}} \,\,\,\,\,\, {\rm for \,\, large} \,\,\,\, k^2 
\ee
(see (\ref{action}, \ref{barh2h0}, \ref{propgauge})).
However, the $n$-graviton interaction has the 
same leading scaling of the kinetic term, since it can be written in the following schematic way,
\be
{\mathcal L}^{(n)} \sim  h^n \, \Box_{\eta} h \,\,  h_i( - \Box_{\Lambda}) \,\, \Box_{\eta} h 
\,\, \rightarrow \,\, h^n \, \Box_{\eta} h 
\,  ( \Box_{\eta} + h^m \, \partial h \partial )^{\gamma + \mathrm{N} } \, 
\Box_{\eta} h , 
\label{intera3}
\ee
where the indexes for the graviton fluctuation $h_{\mu \nu}$ are omitted 
and $h_i( - \Box_{\Lambda})$ is the entire function defined by the properties (i)-(iii). 
From (\ref{intera3}), the upper bound to the superficial degree of divergence 
in a spacetime of ``even" dimension is 
\be
&&\hspace{-0.7cm} 
\delta_{\rm even} = D_{\rm even} L - (2 \gamma + 2 \mathrm{N}+ 4) I + (2 \gamma + 2 \mathrm{N} + 4) V 
= D_{\rm even} L - (2 \gamma + D_{\rm even}) I + (2 \gamma + D_{\rm even}) V \nonumber \\ 
&& \hspace{0.15cm} 
= D_{\rm even} - 2 \gamma  (L - 1).
\label{diverE}
\ee
On the other hand, in a spacetime of ``odd" dimension the upper limit to the degree of divergence is
\be
\delta_{\rm odd} = D_{\rm odd} - (2 \gamma+1)  (L - 1).
\label{diverodd}
\ee
In (\ref{diverE}) and (\ref{diverodd}) we used again the topological relation between vertexes $V$, internal lines $I$ and 
number of loops $L$: $I = V + L -1$. 
Thus, if $\gamma > D_{\rm even}/2$ or $\gamma > (D_{\rm odd}-1)/2$, 
only 1-loop divergences survive in this theory. Therefore, 
the theory is super-renormalizable, unitary and microcausal as pointed out in 
\cite{Krasnikov, efimov, E2, E3, E4, E5}. 
%
Only a finite number of constants is renormalized in the action (\ref{action}), i.e. 
$\kappa$, $\bar{\lambda}$, $a_n$, $b_n$ together with the finite number of couplings that multiply the operators $O(R^3)$ in 
the last line of (\ref{action}). 
The renormalized action reads
{\small
\be 
\hspace{0.7cm}
\boxed{
\begin{array}{rcl}
&&
 \hspace{-0.8cm} \! S= \int d^D x \sqrt{|g|} \Big[2\, Z_{\kappa} \, \kappa^{-2} \, R + Z_{\bar{\lambda}} \bar{\lambda}
+ \sum_{n=0}^{\mathrm{N} } \Big( 
Z_{a_n} \, a_n \, R \, (-\Box_{\Lambda})^n \, R  + 
Z_{b_n} \, b_n \, R_{M N} \, (-\Box_{\Lambda})^n \, R^{M N} 
\Big) 
\\
&& \hspace{0.7cm} 
+ R  \, h_0( - \Box_{\Lambda}) \, R +
R_{M N} \, h_2( - \Box_{\Lambda}) \, R^{M N}
+ 
Z_{c_1^{(1)}} c_1^{(1)} \, R^3 +  \dots\dots\dots + Z_{c_1^{(\mathrm{N})}} \, c_1^{(\mathrm{N})} \, R^{\mathrm{N} +2}
 \, \Big] \label{actionRen}
 \end{array}
 \!\!\!\!   }
\ee 
}

All the couplings in (\ref{actionRen}) must be understood as renormalized at an energy scale $\mu$. 
Contrarily,   
the functions $h_i(z)$ are not renormalized. 
%
We can write the entire functions as a series, i.e. $h_i(z) = \sum_{r=0}^{+\infty} a_r z^r$.
Because of the superficial degrees of divergence (\ref{diverE}) and (\ref{diverodd}), 
there are no counterterms that renormalize $a_r$ for $r >  \mathrm{N}$. 
 Only the 
coefficients $a_r$ for $r \leqslant \mathrm{N}$ could be renormalized; however, 
the other couplings in the first line of (\ref{actionRen}) already incorporate such renormalization.
Therefore, the non-trivial dependence of the entire functions $h_i(z)$ on their argument is preserved 
at quantum level.

Imposing the conditions (i)-(iii) we have the freedom to choose the following form for the functions 
$h_i$,
\be
h_2(z) = \frac{ V(z)^{-1} -1 - \frac{\kappa^2 \Lambda^2}{2} \, z \sum_{n=0}^{\mathrm{N}} \tilde{b}_n \, z^n}{\frac{\kappa^2 \Lambda^2}{2}\, z} \, , \,\,\,\,\,  
h_0(z) = - \frac{V(z)^{-1} -1 + \kappa^2 \Lambda^2 \, z \sum_{n=0}^{\mathrm{N}} \tilde{a}_n \, z^n}{
\kappa^2 \Lambda^2 \, z} \, , 
\label{hzD}
\ee
for general parameters $\tilde{a}_n$ and $\tilde{b}_n$. 
Here $V(z)^{-1}= \exp {H(z)}$ and $H(z)$ 
is an entire function that exhibits logarithmic asymptotic behavior in the conical region $C$. 
The form factor 
$\exp H(z)$ has no zeros in the entire complex plane for $|z|< + \infty$. 
Furthermore, the non-locality in the action is actually a ``soft" form of non locality, because 
a Taylor expansion of $h_i(z)$ eliminates the denominator $\Box_{\Lambda}$. 

An explicit example of $\exp H(z)$ that satisfies the properties (i)-(iii) can be easily constructed \cite{Tombo}. 
The entire function $H(z)$, which is compatible with the property (iii), 
can be defined as 
\be
H(z) = \int_0^{p_{\gamma + \mathrm{N} + 1 }(z)} \frac{1 - \zeta(\omega)}{\omega} {\rm d} \omega \, , 
\label{Hz}
\ee
where $p_{\gamma +\mathrm{N}+1}(z)$ and $\zeta(z)$ must 
satisfy the following requirements:
\begin{enumerate}
\renewcommand{\theenumi}{\alph{enumi}.}
\renewcommand{\labelenumi}{\theenumi}
\item $p_{\gamma +\mathrm{N}+1 }(z)$ is a real polynomial of degree $\gamma+ \mathrm{N} +1$ 
with $p_{\gamma + \mathrm{N}+1}(0) = 0$,
\item $\zeta(z)$ is an entire and real function on the real axis with $\zeta(0) = 1$,
\item $|\zeta(z)| \rightarrow 0$ for $|z| \rightarrow \infty$ in the conical region $C$ defined in (iii). 
\end{enumerate}
There are of course many ways to choose $\zeta(z)$, but we focus here on the exponential choice  
$\zeta(z) = \exp(- z^2)$, which satisfies requirement c. 
in a conical region $C$ with 
$\Theta =\pi/4$. 
The entire function $H(z)$ is the result of the integral (\ref{Hz})
\be
H(z) = \sum_{n =1}^{+ \infty} \, \frac{p_{\gamma +\mathrm{N}+1}(z)^{2 n}}{2n \, ( -1 )^{n-1} \, n!}
= \frac{1}{2} \left[ \gamma_E + 
\Gamma \left(0, p_{\gamma+\mathrm{N}+1}^{2}(z) \right)  + \log \left( p^2_{\gamma+ \mathrm{N}+1}(z) \right) \right] ,
\,\, 
{\rm Re}( p_{\gamma+\mathrm{N}+1}^{2}(z) ) > 0 , 
\label{HD}
\ee
where $\gamma_E=0.577216$ is the Euler's constant and  
$\Gamma(a,z) = \int_z^{+ \infty} t^{a -1} e^{-t} d t$ is the incomplete gamma function.  
If we choose $p_{\gamma+\mathrm{N}+1}(z) = z^{\gamma +N+ 1}$, $H(z)$ simplifies to:
\be
&& 
H(z) = \frac{1}{2} \left[ \gamma_E + \Gamma \left(0, z^{2 \gamma +2 \mathrm{N}+2} \right) + \log (z^{2\gamma 
+2 \mathrm{N}+2}) \right] \, ,
\nonumber \\
&& H(z) = \frac{ z^{2 \gamma + 2 \mathrm{N}+2}}{2} - \frac{ z^{4 \gamma + 4 \mathrm{N}+ 4}}{8} + \dots \,\,\, {\rm for} \,\, z \approx 0  , 
\,\, 
{\rm Re}(z^{2 \gamma +2 \mathrm{N}+2}) > 0. 
\label{H0}
\ee
For $p_{\gamma + \mathrm{N}+1}(z) = z^{\gamma + \mathrm{N}+1}$, the $\Theta$ angle defining 
the cone $C$ is $\Theta = \pi/(4 \gamma +4 \mathrm{N} + 4)$. 
The first correction to the form factor $V^{-1}(z)$ goes to zero faster than any polynomial function for $z \rightarrow + \infty$, namely 
\be
\lim_{z \rightarrow +\infty} V(z)^{-1} = e^{\frac{\gamma_E}{2}} \, |z|^{\gamma + \mathrm{N} +1} \,\,\,\, 
\,\,\,\, {\rm and} \,\,\,\, \,\,\,
\lim_{z \rightarrow +\infty} 
\left(\frac{V(z)^{-1}}{e^{\frac{\gamma_E}{2}} |z|^{\gamma + \mathrm{N} +1} } - 1 \right) z^n = 0
\,\,\,\, \forall \, n \in \mathbb{N}\, .
\label{property}
\ee
The entire function above is completely equivalent to the following one (see also Appendix C),
\be
V^{-1}(z) = e^{\frac{1}{2} \left[ \Gamma \left(0, p_{\gamma + \mathrm{N} + 1}^2(z) \right)+\gamma_E  \right] } \, \left| p_{\gamma + \mathrm{N} + 1} \right|  \, ,
\label{Vlimit1}
\ee
and in the limit of large $z$ the exponential factor goes to one faster than any polynomial, namely
\be
\lim_{z \rightarrow +\infty} e^{\frac{1}{2} \left[ \Gamma \left(0, p_{\gamma + \mathrm{N} + 1 }^2(z) \right)+\gamma_E  \right] } = e^{\frac{1}{2} \gamma_E}.
\label{Vlimit2}
\ee
The property above 
clarifies that 
the ultraviolet corrections to the leading behavior of the form factor do not affect
 the divergent part of the $1$-loop effective action generating instead a convergent integral. 
%
%
For the sake of completeness, we briefly mention 
another possible choice of entire function inspired by string field theory and/or non-commutative effective quantum field theory: it is $H(z) =z^n$ \cite{Krasnikov}. This form factor will be studied more in depth in
another paper about M-theory.

Having obtained an explicit representation of the entire function $H(z)$, we can now expand 
on the power counting renormalizability and on the potential counter terms. 
For this task we will make repeated use of the massless integrals in the Appendix A. 
A general 
one-loop amplitude 
with $n$ external gravitons of momentum $k_i$ and polarizations $e_i$ ($i=1, \dots , n$) reads 
\be
%
A^{(L=1)} \sim \int d^D \ell \frac{ \mathcal{V}_1(\ell_1, k_1, e_1) \,  \mathcal{V}_2(\ell_2, k_2, e_2)  \dots 
 \mathcal{V}_n(\ell_n, k_n, e_n)  }{\ell_1^2 \, V^{-1}(\ell_1) \, \ell_2^2 \, V^{-1}(\ell_2)   
\dots  \, \ell_n^2 \, V^{-1}(\ell_n) } \,  ,
\label{1loopG}
\ee 
where $\ell_i^2 \equiv (\ell - k_1 - k_2 - \dots - k_i)^2$, $\ell_i^2 V^{-1}(\ell_i)$ are the propagator along the loops \cite{amplicompact} and $\mathcal{V}_i(\ell_i, k_i, e_i)$ are vertex operators. We can collect the vertexes in two different sets that may involve 
or not the entire functions $h_i( z)$. Omitting the indexes, we have
\be
&& \hspace{-1.3cm} {\rm set} \, 1  :  \,\,\, R, \, R^2 , \, R^3 , \dots , \, R^{\mathrm{N}+2} \,\,\,   \Longrightarrow \,\,\,\, h^n (\partial^2 h) , \, h^n (\partial^2 h)^2 , \,
h^n (\partial^2 h)^3 , \dots , h^n (\partial^2 h)^{\mathrm{N}+2} \, ,
\nonumber   \\
&& \hspace{-1.3cm} {\rm set} \, 2  : \,\,\, R_{\dots} \, h_i( - \Box_{\Lambda}) R_{\dots} \,\,\,
\Longrightarrow \,\,\,\, h^n (\partial^2 h) h_i(- \Box_{\Lambda}) \, h^m \, (\partial^2 h) = h^n (\partial^2 h) 
\! 
\left[ 
\sum_{r=0}^{+ \infty} c_r (- \Box_{\Lambda})^r 
\right]  \! h^m \, (\partial^2 h) \,  ,
\label{Vertex}
\ee 
where $(i =0,2)$, while ``$n$" and ``$m$" exponents come from the action expansion in the graviton field. 
Given $p_{\gamma+1} = z^{\gamma +{\mathrm N} + 1}$,
if all the vertexes but one come from set.2 in (\ref{Vertex}), then the integral (\ref{1loopG}) 
does not give any logarithmic divergence. 
We find logarithmic divergences only when {\em all} the vertexes come from set.2 in (\ref{Vertex}) and 
the contribution 
to the amplitude is  
\be
&& {\rm divergence} \, : \,\,\,\, k^{2 \mathrm{N} + 4}\int \! d^D \ell \, \frac{1}{\ell^{ \mathrm{N} +2} (\ell - k)^{\mathrm{N} +2 }} \sim \frac{1}{\epsilon} k^{2 \mathrm{N} + 4}
\, , \label{integralg}\\ 
&&{\rm counterterms}\, :  \,\,\,\, \frac{1}{\epsilon} \, R_{\dots}^{\mathrm{N} +2} \,  , \,\,\,\,  \frac{1}{\epsilon} R_{\dots} \Box^{\mathrm{N}} R_{\dots} \,  , 
\,\,\,\,   \frac{1}{\epsilon} R_{\dots} R_{\dots} \Box^{\mathrm{N}-1} R_{\dots}    \,\,\, , \dots \,  ,
\label{ellediv}
\ee
where we used the asymptotic property (\ref{property}) and $\epsilon$ is the ultraviolet cut-off in dimensional 
regularization. 
If more momentum factors are attached to internal lines, then the loop amplitude (\ref{1loopG}) leads to the same 
counterterms as those in (\ref{ellediv}) (this result follows from the integral (\ref{I0}) in the Appendix A), or to zero
(this follows from the integrals (\ref{Ikn}) and (\ref{null}) in Appendix A.) 
On the other hand, 
if more momentum factors are attached to external lines, then the loop integral is convergent. 
The outcome is that, for a large enough $\gamma$, we have counterterms only at the order $R^{\mathrm{N} +2}$.
For example, in $D=4$ the counterterms are $R^2$ and $R_{\mu \nu}^2$, but there are no divergent contributions 
proportional to $R$ or $\bar{\lambda}$ (cosmological constant). 
This is a property of the theories defined in (\ref{H0}) by the particular polynomial $p_{\gamma+\mathrm{N}+1}(z) = z^{\gamma +N+ 1}$. If we instead consider a more general polynomial such as  
\be
p_{\gamma+\mathrm{N}+1}(z) = a_{ 1} z^{\gamma +N+ 1} + a_{0} z^{\gamma +N} + a_{-1}z^{\gamma +N-1} + \dots + z \label{poly} \,  ,
\ee
 then the other couplings, concerning the counterterms with less derivatives, 
 are renormalized
as explicitly stated in (\ref{actionRen}).

\subsection{Renormalization \& asymptotic freedom}
We are now ready to expand on the renormalized Lagrangian in (\ref{actionRen}).
We start with the classical action written in terms of renormalized couplings, and then we add counterterms to subtract divergences.
The counterterms may be displayed by explicitly adding and subtracting the classical action in 
$\mathcal{L}_{\rm ren}$ (\ref{actionRen}),
\be 
&&
 \hspace{-0.7cm} 
 \mathcal{L}_{\rm ren} = \mathcal{L} + \mathcal{L}_{\rm ct}   
 = \mathcal{L} + 
2   (Z_{\kappa} -1)  \kappa^{-2} \, R + (Z_{\bar{\lambda}} -1) \bar{\lambda} 
+ (Z_{c_1^{(1)}} -1 )c_1^{(1)}  R^3 +  \dots  + \, (Z_{c_1^{(\mathrm{N})}}-1)c_1^{(\mathrm{N})}  R^{\mathrm{N} +2} 
  \nonumber \\
&& \hspace{2.39cm}
+ \sum_{n=0}^{\mathrm{N} } \Big( 
 \,  ( Z_{a_n} -1)a_n  \, R \, (-\Box_{\Lambda})^n \, R  + 
 \, ( Z_{b_n} -1)  b_n \, R_{M N} \, (-\Box_{\Lambda})^n \, R^{M N}   \Big) \,  , 
 \label{actionRen2}
\ee 
where $\mathcal{L}_{\rm ct}$ is Lagrangian of the counterterms. In dimensional regularization, the latter  Lagrangian 
looks like
{\small
\be
\hspace{0.1cm} 
 \mathcal{L}_{\rm ct} 
 = \frac{1}{\epsilon} \left[ \beta_{\kappa} R + \beta_{\bar{\lambda}}
 + \sum_{n=0}^{\mathrm{N} } \, \Big( \beta_{a_n} R \, (-\Box_{\Lambda})^n \, R  + 
  \beta_{b_n}  \, R_{M N} \, (-\Box_{\Lambda})^n \, R^{M N}   \Big) 
 +
\beta_{c_1^{(1)}}  \, R^3 +  \cdots + \, \beta_{c_1^{(\mathrm{N})}} 
\, R^{\mathrm{N} +2} \right] \! , \label{Abeta2}
\ee
}
\hspace{-0.3cm}
where $$\beta_{\kappa}, \beta_{\bar{\lambda}}, \beta_{a_n}, \beta_{b_n}, \beta_{c_1^{(1)}}, \cdots, \beta_{c_1^{(\mathrm{N})}}$$ are the beta functions of the theory and $\epsilon$ is the ultraviolet cutoff.
Since the one-loop Green functions obtained from the 
effective action must be finite 
when $\epsilon \rightarrow 0$,  
the counterterms Lagrangian is related to the divergent part of the effective Lagrangian by 
$\mathcal{L}_{\rm ct} = -  \mathcal{L}_{\rm div}$. 
The effective action and the beta functions can be calculated using the techniques developed by 
Barvinsky and Vilkovisky in \cite{BV} (see Appendix B). 
Comparing (\ref{actionRen2}) and (\ref{Abeta}), we find 
\be
 (Z_{\alpha_i} -1) \alpha_i = \frac{1}{\epsilon} \beta_{\alpha_i} \,\,\, \Longrightarrow \,\,\, 
 Z_{\alpha_i} = 1 + \frac{1}{\epsilon} \beta_{\alpha_i} \frac{1}{\alpha_i} \, ,
\ee
where $\alpha_i$ is one of the coupling constants in the action, 
\be
\alpha_i  \in \{\kappa, \bar{\lambda}, a_n, b_n, c_1^{(1)}, \cdots, c_1^{(\mathrm{N})} \} \equiv 
\{\kappa, \bar{\lambda}, \tilde{\alpha}_n \} 
\label{couplings}
 \, .
\ee 
The bare $\alpha_i^{B}$ and the renormalized $\alpha_i$ coupling constants come together in  
$\alpha_i^{B} = \alpha_i \, Z_{\alpha_i}$, so the running of $\alpha_i(\mu)$ with the energy scale 
``$\mu$" in the ultraviolet regime is 
\be
\alpha_i(\mu) \sim \alpha_i(\mu_0) + \beta_i \log \left( \frac{\mu}{\mu_0} \right)^{ Y_i} , 
\label{alphasol}
\ee
where the exponents $Y_i$ can be obtained by solving exactly the renormalization group equations
in the ultraviolet regime:
\be
\frac{d \alpha_i}{d t} = \beta_i(\alpha_i) \, , \,\,\,\,\,  t = \log \left( \frac{\mu}{\mu_0} \right) . 
\ee
If all the $\beta_i$ functions flow to a constant in the ultraviolet regime, then the exponents $Y_i=1$. This is exactly the case of
our theory because no divergence comes from the vertexes in set 1 (\ref{Vertex}) and the beta functions 
result independent from the coupling constants $\alpha_i$ (see Appendix B). We can conclude 
that $\alpha_i(\mu) \sim \alpha_i(\mu_0) + \beta_i \, t$, which proves that the theory is {\em asymptotically free} \cite{TomboAF}. 
In this theory, the 
sign of the beta functions and the 
sign of the logarithmic finite contributions to the one loop effective action do not play any crucial role, because the 
leading asymptotic behavior of the dressed propagator is entirely due to its bare part.
The self energy insertions do not contribute to this process and therefore unitarity is preserved
(see next section).


\subsection{Unitarity}
Let us investigate the unitarity of the theory.
We assume that the theory is renormalized at some scale $\mu_0$; therefore, 
if we set  
\be
 \tilde{a}_n = a_n(\mu_0) \,\, , 
 \,\,\,\, \tilde{b}_n = b_n(\mu_0),
\label{betaalphaD}
\ee
the bare propagator does not possess other gauge-invariant pole than 
the physical graviton one and 
\be
\bar{h}_2 = \bar{h}_0 = V(z)^{-1} = \exp H(z).
\ee
%
%
Thus, only the physical massless spin-2 graviton pole
occurs in the bare propagator and (\ref{propgauge}) reads 
\be 
 \mathcal{O}^{-1}(k)^{\xi = 0} = \frac{V(k^2/\Lambda^2)  } {k^2}
\left( P^{(2)} 
- \frac{P^{(0)}}{D-2 }  \right).
\label{propgauge2}
\ee
The momentum or energy scale at which the relation between the quantity computed and the quantity 
measured is identified is called the subtraction point and is indicated usually by ``$\mu$". 
The subtraction point is arbitrary and unphysical, so the final answers do not have to depend 
on the subtraction scale $\mu$. Therefore, the derivative $d/d \mu^2$ of physical quantities has to be zero.
In our case, if we choose another renormalization scale $\mu$, then the bare propagator acquires poles. 
However, these poles 
cancel out in the dressed physical propagator because the shift in the bare part is cancelled 
by a corresponding shift in the self energy. 
The renormalized action (\ref{actionRen}) will produce finite Green's functions to whatever order 
in the coupling constants we have renormalized the theory to. For example, the $2$-point Green's 
function for the spin $2$ and spin $0$ sectors at the first order 
in the couplings $a_n$, $b_n$ 
can be schematically written as
\be
[\mathcal{O}^{-1}_{R}]^{-1}(k) \sim V^{-1} \left( k^2/\Lambda^2 \right) \, (k^2 + \Sigma_R(k^2)),
\label{qprop}
\ee
where the renormalization prescription requires that $\Sigma_R$ satisfies (on shell)
\be
\Sigma_R(0) = 0 \,\,\,\,\,\, {\rm and} \,\,\,\,\,\, \frac{\partial \Sigma_R}{\partial k^2} \Big|_{k^2 = 0} =0.
\ee
According to the power counting analysis, in four dimensions we have 
\be
\Sigma_R(k^2) \sim 
\kappa^2 \, V(k^2/\Lambda^2) \, k^4 \,  \log \left( \frac{k^2}{\mu^2} \right) \,\,\, \,\, 
\Longrightarrow \,\,\, \,\, 
\frac{V \left( k^2/\Lambda^2 \right)}{k^2\left[ 1 +  \, c_0\, V(k^2/\Lambda^2)   \, k^2 \,  \log \left( \frac{k^2}{\mu^2} \right)\right] },
\ee
while in $D$ dimensions we expect  
\be
&& \Sigma_R(k^2) \sim 
V\left( k^2/\Lambda^2 \right) 
\left[ c_0 k^4 + \dots + c_{\rm N} k^{2 \mathrm{N} + 2} \,
\right]   
\log \left( \frac{k^2}{\mu^2} \right) \, , \label{SIGMA} \\
&& \mathcal{O}^{-1}_{R}(k) \sim
\frac{V \left( k^2/\Lambda^2 \right)}{k^2\left\{ 1 + V(k^2/\Lambda^2) \, \left[ c_0 k^2 + \dots + c_{\rm N} k^{2 \mathrm{N} + 2} \,
\right]  
\,  \log \left( \frac{k^2}{\mu^2} \right)\right\}} \, . 
\ee
The subtraction point is arbitrary and therefore we can take the renormalization prescription 
off-shell to $k^2 = \mu^2$. The couplings we wish to renormalize must be dependent 
on the chosen subtraction point,
$a_n(\mu)$ and $b_n(\mu)$, in such a way that the experimentally 
measured couplings do not vary on-shell.   
The renormalized Green's function $[\mathcal{O}^{-1}_{R}]^{-1}(k)$ at $\mu^2$
must produce the same 
Green's function when expressed in terms of bare quantities.
Consequently, 
the scalings $Z_{a_n}$ and $Z_{b_n}$ must also depend on  
$\mu^2$. 
The Green's function written in terms 
of bare quantities can not depend on $\mu^2$, but since $\mu^2$ is arbitrary, the renormalized Green's 
function must not depend either. This fact, $\partial_{\mu^2} \, \mathcal{O}^{-1}_{R}(k)  = 0$ 
is known as the renormalization group invariance.
In addition to what has been said up to now,
an analysis of the amplitudes shows that unitarity 
can be satisfied because the propagator (\ref{propgauge2}) is simply the 
Einstein's theory propagator multiplied by an entire function \cite{Tombo}. Thus, the positions of the singularities in the finite region of the complex momentum plane are unaltered and the Cutkosky cutting rules can be satisfied.
The quantum correction $\Sigma_R(k^2)$ given in (\ref{SIGMA}) 
goes to zero in the ultraviolet regime, i.e.
\be
 \lim_{k^2 \rightarrow + \infty} \Sigma_R(k^2) = 0 \, ,
\ee
and 
the leading asymptotic behavior of the dressed propagator is entirely due to its bare
component, i.e. self-energy insertions do not contribute to it \cite{Tombo}. 

We conclude this section presenting the classical equations of motion.
Using 
(\ref{betaalphaD}) the 
bare Lagrangian (\ref{action}) at the chosen scale $\mu_0$ 
 simplifies to
\be
\hspace{0.4cm}
\begin{centering}
 \boxed{
\begin{array}{rcl}
 \mathcal{L} = \sqrt{|g|} \Big\{2 \kappa^{-2} \Big[ R  
- G_{M N} \, \gamma(\Box) \,  R^{M N} \Big]   
+\underbrace{ R^3 + \dots + R^{D/2} }_{{\rm local} \,\, {\rm operators }}
\Big\} \, , \,\,\,\, 
\gamma(\Box) :=  \frac{ V(-\Box_{\Lambda})^{-1} -1}{\Box} 
\end{array}
}
\end{centering}
\label{ActionD}
\ee
The equations of motion at the order $O(R^2)$ reads 
\be
e^{H(-\Box_{\Lambda})} \, G_{M N} + O(R^2) = 8 \pi G_N \,  T_{M N} \, . 
\label{EMT}
\ee
Furthermore, we point out that the first correction to the equations of motion is $O(R^2)$ and no other derivative terms 
that are linear in the curvature appear.
The exact equations of motion are more convoluted and they can be obtained 
by applying the result in \cite{koshe1} to action (\ref{ActionD}). 
The truncated equations (\ref{EMT}) have been solved for a spherically symmetric spacetime, 
highlighting the absence of singularities in the entire spacetime \cite{MMN, NicoNew, NS}.

\subsection{Coincidence limit in the two-point function}
In this section, we are going to  
evaluate the coincidence limit in the two-point function \cite{efimov}. For this evaluation 
we should first determine the 
propagator in the coordinate space. 
For a general 
form factor $V(z)$ the Fourier transform of (\ref{propgauge2}) reads 
\be
\hspace{0cm} \tilde{\mathcal{O}}^{-1}(x) = \int \frac{d^D k}{(2 \pi)^D} \, \frac{V(k^2 \, \ell^2)}{k^2} \, e^{ i k x}  \,  , \,\,\,\, \ell \equiv1/\Lambda \, ,
\label{Gx}
\ee
where we neglected any tensorial structure and we assumed Euclidean signature. 
Changing the existing coordinates into $D$-dimensional spherical ones and integrating 
(\ref{Gx}) in 
the angular variables, we get 
\be
\hspace{0cm}
\tilde{ \mathcal{O}}^{-1}(x)  = \frac{\pi^{\frac{D-3}{2}}}{(2 \pi)^{D-1} \,  \Gamma\left( \frac{D-1}{2} \right)} 
\int_0^{+\infty} d u \, \frac{u^{\frac{D-4}{2}} V(u \, \ell^2) }{2} 
 \sqrt{\pi} \, \Gamma\left( \frac{D-1}{2} \right) \! \,_0\tilde{F}_1\!\!\left( \frac{D}{2}; - \frac{u \, x^2}{4} \right) ,
\label{propcoord}
\ee
where we have introduced the variable $u = k^2$. 
For $u \rightarrow + \infty$, $V(u \, l^2) \sim 1/ u^{\gamma + \mathrm{N}+1}$, while for $x^2 \rightarrow 0$, $\,_0\tilde{F}_1\approx {\rm const.}$, the propagator in the coincidence limit is finite only for certain values of $\gamma$,
\be \hspace{-0.5cm} 
\tilde{\mathcal{O}}^{-1}(0)  \propto \int_0^{+\infty} d u \,  u^{ \frac{D-4}{2} - (\gamma + \mathrm{N}+1)} < \infty \,\,\, 
\Longleftrightarrow \,\, 2 \gamma + 2 \mathrm{N} +4  > D.
\ee

\subsection{
 Multidimensional renormalizable gravity?}
\noindent
Relying on the action introduced in (\ref{action}),
a potential renormalizable and unitary theory
can be defined by the following 
form factor, 
\be
&& V(z) = e^{-H(z)} \, , \label{modefactor}\\
&& H(z) = \frac{1}{2} \left\{ \gamma_E + 
\Gamma \left(0, z^{2 \mathrm{N}+2}\right)  + \log [ z^{2 \mathrm{N}+2} ] \right\} , 
\nonumber \\
&& {\rm Re}( \, z^{2 \mathrm{N}+2} ) > 0.
\nonumber 
\ee
The form factor (\ref{modefactor}) has been developed from (\ref{HD}) by choosing $\gamma = 0$
and the behavior of the entire functions $h_i(z)$ for $|z| \rightarrow + \infty$ is 
\be
&&\hspace{-0.2cm}
 \lim_{|z|\rightarrow + \infty} |h_i(z)| \rightarrow | z |^{\mathrm{N}} \, , \,\,\,\, 
\label{hire}
{\rm for} \,\, z  \,\, {\rm in} \,\, {\rm the} \,\, {\rm cone}: \\
&& \hspace{-0.2cm} 
C = \{ z \, | \,\, - \Theta < {\rm arg} z < + \Theta \, , \,\,  \pi - \Theta < {\rm arg} z < \pi + \Theta \} \, , 
\,\,\,\, 
{\rm for } \,\,\, 
\Theta = \pi/(4 \mathrm{N} +4) \, .
\nonumber 
\ee
%
For the theory in question 
the amplitudes are divergent at each order in the loop expansion and 
the maximal superficial degree of divergence from (\ref{deltaD}) or (\ref{DDX})  
is $\delta = D$ as it occurs in the local theory.  Therefore, the theory ceases to be super-renormalizable, 
and at a first glance it seems to preserve  renormalizability and unitarity as it can be inferred from (\ref{propgauge2}) 
with the entire function $H(z)$ defined in (\ref{modefactor}). 
In four dimensions $\mathrm{N}=0$ and the theory flows to the Stelle's higher derivative gravity in the ultraviolet regime.  
Regrettably, a more careful analysis 
highlights the need to introduce nonlocal counter terms that renormalize the entire functions.
This disproves our main hypothesis of non-renormalization of the two non-polynomial functions $h_2(z)$ and $h_0(z)$ \cite{Tombo}. However, this theory may be helpful as a toy model 
to hypothesize the asymptotic freedom in all the super-renormalizable theories here introduced. 

%


\section{Finite quantum gravity} 
\label{secFQG}
In this section we emphasize that in odd dimension the pure gravitational theory is finite at quantum level.
We have already showed in the previous sections that the non-polynomial extension of Einstein theory
provides a super-renormalizable theory with only one loop divergences. 
However, in {\em odd} dimension there are no local invariants (in dimensional regularization) with an odd number of derivatives which could serve as counter-terms for pure gravity. 
This is a consequence of the rational nature of the entire functions which characterize the theory (one example of non rational function is $h_i(\sqrt{z})$.)  
Therefore, following the Duff analysis \cite{duff}, quantum gravity is finite in even dimension, as well, once the Kaluza-Klein compactification is 
applied. The finiteness of the theory in even dimensions follows from the inclusion 
of an infinity tower of states which drastically affects the ultraviolet behavior. 
When matter fields are added, the one-loop finiteness of the theory in odd dimension could be spoiled, but this does not happen
to a super-symmetric extension of the theory, which remains finite \cite{duff}. 
%
Even though the super-symmetric extension of a non-polynomial action is not going to be
tackled in this paper, 
a preliminary analysis has been done for $N=1$ supergravity in four dimensions and $N=1$, $N=2$ 
supergravity in three dimensions \cite{NLsugra}. In \cite{NLsugra} supersymmetry was realized
for the non-polynomial higher derivative gravity 
filling up the supergravity multiplets carefully avoiding extra poles in the propagators \cite{FerraraExt}.  

Let us 
expand on the finiteness of the pure gravitational theory in odd dimension. 
Given a compact $D$-dimensional manifold $\mathcal{M}$, the most efficient way for discussing renormalizability and
computing counterterms is to evaluate the effective action using the background field method and the heat kernel
techniques.
The typical one-loop effective action arises from performing the functional integral over the classical action,
which is quadratic in the fields and is given by 
\be
W = \frac{1}{2} \log \det F(\Delta)=   \frac{1}{2} {\rm Tr} \log  F(\Delta)=
- \frac{1}{2} \int_0^{+ \infty} \frac{dt }{t} \, {\rm Tr} \left( e^{ - F(\Delta) t} \right)  
 \, ,
\label{effaction}
\ee
where $F(\Delta)$ is the operator representing the quadratic part of the full action including gauge fixing and FP-ghost action. 
The background field method implies the following 
special parametrization of the metric
\be
g_{M N} \rightarrow g_{M N}^{\prime} = {g}_{M N} + h_{M N},
\ee 
where $g_{M N}$ is the background metric and $h_{M N}$ is the quantum field. 
In (\ref{effaction}) we introduced a parameter $t$, called fictitious time, and we defined the traced ``heat kernel"
\be
K(\Delta,t) = {\rm Tr} \left( e^{ - F(\Delta) t } \right) = \int d^D x \sqrt{g} \, K(\Delta; x,x;t) . 
\label{KDelta}
\ee
%
It is not necessary to choose a particular entire function $H(z)$ 
because the argument that follows hereby is general and independent from the particular action if the properties listed at the end of section (\ref{gravitonpropagator})  are satisfied. 
In this section, the locality of the 
counterterms, which is guaranteed by the super-renormalizability of the theory, will be crucial. 

In the limit $t \rightarrow 0^+$ it can be shown that $K(\Delta,t)$ manifests the following asymptotic expansion 
\be
K(\Delta, t) =\sum_{k=0}^{+\infty} t^{-\frac{(D-k)}{2}} \, A_k(\Delta) \, , \,\,\,\, 
A_k(\Delta) = \frac{1}{( 4 \pi)^{\frac{D}{2}}} \int d^D x \, a_k(\Delta) \, ,
\label{Kt0}
\ee
where $a_k(\Delta)$ are $D$ independent invariants constructed with $g_{M N}$ and its derivatives  
of order $k$, 
\be
a_0 = 1 \, , \,\,\, a_2 = a_2^{(1)} R \, , \,\,\, a_4 = a_4^{(1)} R_{M N P Q}^2 +a_4^{(2)} R_{M N}^2 +
a_4^{(3)} R^2 \, , \,\,\, \dots \, .
\label{coeff}
\ee
When we substitute the expansion (\ref{Kt0}) in 
the effective action (\ref{effaction}), we find that it is made of both convergent and divergent contributions, 
\be
W = 
- \frac{1}{2}  \int_0^{t_0}   \, \sum_{k=0}^{+ \infty} \,  \frac{dt}{t} \, t^{-\frac{(D-k)}{2}} \, A_k(\Delta) 
- \frac{1}{2}  \int_{t_0}^{+ \infty }  \, \sum_{k=0}^{+ \infty} \,  \frac{dt}{t} \, t^{-\frac{(D-k)}{2}} \, A_k(\Delta)
\,\,\,\,\,  ( 0 < t_0 \leqslant 1) 
\, .
\label{Wdiv}
\ee
This quantity 
may be regularized adopting dimensional regularization where the spacetime dimension $D$ is treated as a regularizing parameter. 
Equation (\ref{Wdiv}) is analytical for Re$(D)<0$  and,  
replacing $D$ with $D + \epsilon$, 
we have the following divergent contribution 
to the effective action
\be
W_{\rm div} 
= - 
 \sum_{k=0}^{+ \infty} \,  \frac{ t_0^{-\frac{(D+ \epsilon-k)}{2}} \, A_k(\Delta) }{D + \epsilon -k}
 = - \frac{1}{\epsilon} A_D(\Delta). 
 \label{Wdiv2}
\ee
In (\ref{Wdiv2}) only $A_D(\Delta)$ is non zero but in general for an higher derivative theory 
all the even coefficients $a_k$ may occur in the effective action, while 
the constants $a_k^{(n)}$ in (\ref{coeff}) depend on the particular theory under consideration, namely 
\be
&& \hspace{-0.85cm}
W_{\rm div} 
= -   \sum_{k=0}^{+ \infty} \,  \left\{ \frac{ t_0^{-\frac{(D+ \epsilon-k)}{2}} \, c_D \, A_k(\Delta) }{D + \epsilon -k}
 + \frac{ t_0^{-\frac{(D+ \epsilon-k -2)}{2}} \, c_{D-2} \, A_k(\Delta) }{D + \epsilon -k -2 } 
 + \dots +
 \frac{ t_0^{-\frac{(D+ \epsilon-k-D )}{2}} \, c_{0} \,A_k(\Delta) }{ D+\epsilon-k -D}\right\} \nonumber \\
&&
= - \frac{1}{\epsilon} \left[ c_D \, A_D(\Delta) + c_{D-2} \, A_{D-2}(\Delta) + 
c_{D-4} \, A_{D-4}(\Delta) + \dots + c_0 \, A_0(\Delta) \right].
\ee
The constants $c_i$ $(i=D,D-2, D-4, \dots, 0)$ also depend on the particular theory, 
%
%
%
while
the odd coefficients $a_1,a_3,a_5, \dots$ vanish for a manifold $\mathcal{M}$ without boundary 
because there are no local  
invariant operators with an odd number of derivatives for pure gravity in odd dimensions. 
The rational nature of the non-polynomial entire functions 
and the locality of the counterterms are crucial to this purpose. 
This is true for supergravity, as well \cite{duff}. 
We conclude that all 
the amplitudes with an arbitrary number of loops are finite and all the beta functions
are identically zero in odd dimension,
\be
\beta_{a_n} = \beta_{b_n} = \beta_{c_i^{(n)}} = 0\,,  \,\,\,\, i \in \{1, \dots, ({\rm number \,\, of \,\, invariants \,\, of \,\, order} \,\, N)\} \, , \,\,\,\, n =1, \dots, N .
\label{betaf}
\ee
It follows that we can fix to zero all the couplings $c_i^{(n)}$ and set to constants the couplings 
$a_n(\mu)$, 
$b_n(\mu)$, namely 
\be
c_i^{(n)} = {\rm const.} \,\, , \,\,\,\, a_n(\mu) = {\rm const.}= \tilde{a}_n \,\, , \,\,\,\, b_n(\mu) = {\rm const.}= \tilde{b}_n.
\label{costanti}
\ee
Due to the result (\ref{betaf}) and  
using (\ref{hzD}, \ref{betaalphaD}, \ref{costanti}), the Lagrangian for gravity in odd dimension simplifies to 
%
\be
\hspace{0.0cm}
\boxed{
\begin{array}{rcl}
&& 
\mathcal{L}_{\rm odd} = \sqrt{|g|} \, 2 \kappa^{-2} \Big( R  
- G_{M N} \, \gamma(\Box) \,  R^{M N} \Big) 
\, , \,\,\, \, \gamma(\Box) :=  \frac{ V(-\Box_{\Lambda})^{-1} -1}{\Box} \, .
\end{array}
}
\ee

\section{Conclusions}
In this paper, we focused on 
a class of extended theories of gravity of ``non-polynomial" or ``semi-polynomial" nature in multidimensional spacetime. The non-polynomiality of the classical action is embedded in an entire function (form factor) whose role is twofold.
On the one hand, the form factor is able to improve the convergence of the loop amplitudes 
making the theory super-renormalizable in any dimension (only one-loop divergences survive); 
on the other hand, the form factor ensures unitarity avoiding the onset of ghosts and other degrees of freedom in the spectrum. 
Moreover, the tensorial structure of pure gravity enables us to conclude that the theory is {\em finite} in 
{\em odd dimension}, 
because there are no local invariant operators with an odd number of derivatives which could serve as counterterms in odd dimension.  
The action in odd dimension is very simple because we can consistently fix to zero many of the coupling
constants. It is therefore worth repeating once again the structure of the action,
\begin{empheq}[box={\mybluebox[2pt][2pt]}]{equation*}
S = \int \!  d^{D_{\rm odd}} x \, 2 \kappa^{-2} \, \sqrt{|g|} \left[ R - G_{M N} \left(\frac{ e^{H(-\Box_{\Lambda}) } -1 }{\Box} \right) R^{M N} \right]  
\end{empheq}
\begin{empheq}[box={\mybluebox[2pt][2pt]}]{equation*}
e^{H(z)} = \left| p_{\gamma + \mathrm{N} + 1}(z) \right|  \, e^{\frac{1}{2} \left[ \Gamma \left(0, p_{\gamma + \mathrm{N} + 1}^2(z) \right)+\gamma_E  \right] } \, , \,\,\,\, 
p_{\gamma + \mathrm{N} + 1}(z): \,\, {\rm real \,\, polynomial \,\, of \,\, degree} \,\, 
\gamma + \mathrm{N} + 1 \, .
\end{empheq}
%
{\color{cyan}
}

A further comment regarding the non-polynomial or non-local nature of the action is needed. 
People are usually skeptical about non-locality in general, overlooking that all the known interactions 
are characterized by a one-loop non-local effective action \cite{Maggiore}. 
What we did here is just to start out with a non-polynomiality in the classical action. 
Within the quantum field framework, if we want to preserve Lorentz or diffeomorphism invariance and 
unitarity, while at the same time to have a renormalizable theory of gravity, then we are forced 
to introduce at least one entire function in the action. 

Classical solutions of the equation of motion have been already studied in literature showing the regularity of the spacetime
in the ultraviolet regime. The singularities that plague Einstein's gravity
are here smeared out by the behavior of the theory at short distances \cite{koshe1, koshe2, koshe3, koshe4}.



\section{Appendix A. Massless integrals}
The following massless integrals are useful to understand the conclusions about renormalizability and 
counterterms of the gravity theory presented in this paper. 
\be
&& \hspace{-1.4cm}
\mathcal{I}_{0} = \int d^D p \, \frac{1}{p^{2 a} (p-k)^{2 b}} \sim \frac{1}{\epsilon} \, k^{D - 2a -2 b} \, ,
\label{I0}\\
&& \hspace{-1.4cm}
\mathcal{I}_{n-{\rm even}} =  \int d^D p \, \frac{p_{\mu_1} p_{\mu_2} \dots p_{\mu_n}}{p^{2 a} (p-k)^{2 b}} 
\sim \frac{1}{\epsilon} \, \Big\{k^{D - 2a -2 b +n} \, (\eta_{\mu_1 \mu_2} 
\dots \eta_{\mu_{n-1} \mu_n} + {\rm permut.}) +  \nonumber \\
&& \hspace{4.4cm}
k^{D- 2a -2 b+n-2} \, (k_{\mu_1} k_{ \mu_2} \, \eta_{\mu_3 \mu_4} 
\dots \eta_{\mu_{n-1} \mu_n} + {\rm permut.}) + 
\nonumber \\
&& \hspace{4.4cm}
k^{D- 2a -2b} \, k_{\mu_1} k_{ \mu_2} \dots k_{\mu_n} \Big\}  \, \\
&& \hspace{-1.4cm}
\mathcal{I}_{n-{\rm odd}} = \int d^D p \, \frac{p_{\mu_1} p_{\mu_2} \dots p_{\mu_n}}{p^{2 a} (p-k)^{2 b}} 
\sim \frac{1}{\epsilon} \, \Big\{k^{D - 2a -2 b +n-1} \, (k_{\mu_1} \eta_{\mu_2 \mu_3} 
\dots \eta_{\mu_{n-1} \mu_n} + {\rm permut.}) +  \nonumber \\
&& \hspace{4.4cm}
k^{D- 2a -2 b+n-3} \, (k_{\mu_1} k_{ \mu_2} k_{\mu_3}\, \eta_{\mu_4 \mu_5} 
\dots \eta_{\mu_{n-1} \mu_n} + {\rm permut.})  + 
\nonumber 
\\
&& \hspace{4.4cm}
k^{D- 2a -2b} \, k_{\mu_1} k_{ \mu_2} \dots k_{\mu_n}\, 
\Big\} 
, \label{gene}  \\
&& \hspace{-1.4cm}
\mathcal{I}_{k,n}  = 
\int d^D p \frac{(p^2)^k}{(p^2+ C )^n} = i \frac{C^{\frac{D}{2}-(n-k)}}{(4 \pi)^{\frac{D}{2}}} \, \frac{\Gamma(n-k - D/2) \Gamma(k+D/2)}{\Gamma(D/2) \Gamma(n)}  \,  , 
\label{Ikn}\\
&& \hspace{-1.4cm}
\mathcal{I}_{\rm null} = \int d^D p \frac{1}{p^{2 N} } \equiv 0 \,\,\, {\rm for} \,\,\, N < D/2 \, .\label{null}
\ee
Taking the limit $C\rightarrow 0$ and $N= n-k$ in (\ref{Ikn}) the last integral 
(\ref{null}) follows. 
In {\em odd dimension} the integrals $\mathcal{I}_{k,n}$ are finite if the exponents ``$n$" and ``$k$" are integer
numbers (this 
due to the rational nature of the entire functions as explained in section
\ref{secFQG}). 
Since the gamma function $\Gamma(z)$ has poles in 
$\{0,-1, -2, \dots \}$, $\Gamma(n-k - D/2)$ takes finite values in odd dimension because $D/2$ is a 
semi-integer. It follows that there are no logarithmic divergences at one loop in odd dimension. 
Another useful expression to bring all the integrals in momentum space to the form (\ref{Ikn}) is 
\be
\frac{1}{a^{\alpha} b^{\beta}  } = \frac{\Gamma(\alpha + \beta)}{  \Gamma(\alpha) \Gamma(\beta)}
\int_0^1 dx \, \frac{x^{\alpha-1} (1-x)^{\beta-1}}{[ a x + b (1-x)]^{\alpha + \beta}} \, .
\label{fai}
\ee
Using (\ref{fai}) we can expand every integral (\ref{gene}) in a basis of integrals (\ref{Ikn}).

\section{Appendix B. One loop effective action}
Here we review the calculation of the one loop effective action following the previous analysis 
of Asoreya, L$\grave{{\rm o}}$pez and Shapiro \cite{shapiro}, which used of the Barvinsky-Vilkovisky 
\cite{GBV} techniques. 
The full action consists of the gravitational sector (\ref{action}), gauge fixing condition and FP-ghost action,
namely  
\be
&& 
S_{\rm gf}  = \int d^D x \sqrt{|g|} \, \chi_{M} \, C^{M N} \, \chi_{N} \, , \nonumber \\ 
&& \chi_{M} = \nabla_{P} h^{P}_{M} - \beta_g  \nabla_{M} h \, , \nonumber \\
&&C^{M N} = - \frac{1}{\alpha_g} \left( g^{M N} \Box + \gamma_g \nabla^{M} \nabla^{N} - \nabla^{N} \nabla^{M} \right) \left( \frac{\Box}{\Lambda^2} \right)^{{\rm N} + \gamma} \, , \nonumber \\
&& 
S_{\rm gh} = \int d^D x \sqrt{|g|} \, \bar{C}_{\alpha} \, M^{\alpha}_{\beta} \, C^{\beta} \, , \nonumber \\
&& 
M_{\alpha}^{\beta} = \Box \delta_{\alpha}^{\beta} + \nabla_{\alpha} \nabla^{\beta} - 2 \beta_g \nabla^{\beta} \nabla_{\alpha}. 
\ee
We use the covariant generalization of the gauge fixing (\ref{GF2}) with $\lambda_1 := - 1/\alpha_g$,
$\lambda := \beta_g$, $\lambda_2 = \lambda_3 =0$, and 
\be
\omega_1^{M N}( - \Box_{\Lambda}) : = \left( g^{M N} \Box + \gamma_g \nabla^{M} \nabla^{N} - \nabla^{N} \nabla^{M} \right)  \Box_{\Lambda}^{{\rm N} + \gamma}.
\ee 
Given the properties (\ref{Vlimit1}) and (\ref{Vlimit2})(see also Appendix C), 
for $\gamma>D/2$ 
the vertexes in set 1 do not contribute to the one loop divergences which result only
from set 2. 
Therefore, the non-polynomial operators 
essential to calculate the divergent contribution to the one loop effective action read
{\small 
\be
&& \hspace{0cm}
R  \, h_0( - \Box_{\Lambda}) \, R 
\rightarrow 
- \frac{e^{\frac{\gamma_E}{2}}}{\kappa^2 \Lambda^2} \,
R    \left[  a_1 \left( \frac{ -\Box}{\Lambda^2} \right)^{\gamma +{\rm N} } + 
                 a_0 \left( \frac{ - \Box}{\Lambda^2} \right)^{\gamma +{\rm N} -1 }  \!\!\!\! +
                a_{-1} \left( \frac{ - \Box}{\Lambda^2} \right)^{\gamma +{\rm N} -2 } \!\!\!\! + \dots \right] \! R 
                \nonumber \\
                && \hspace{2.7cm}
               \equiv R    \left[   \omega_{{\rm N} } \, \Box^{\gamma +{\rm N} } + 
                 \omega_{ {\rm N} -1}   \, \Box ^{\gamma+{\rm N} -1 }  +
                 \omega_{ {\rm N} -2}  \,  \Box^{\gamma +{\rm N} -2 } + \dots \right]   \! R \, , \nonumber \\
        &&  \hspace{0cm}
        R_{M N} \, h_2( - \Box_{\Lambda}) \, R^{M N}  \rightarrow 
 \frac{2 e^{\frac{\gamma_E}{2}}}{\kappa^2 \Lambda^2} \, 
R_{M N} \!
 \left[  a_1 \left( \frac{ -\Box}{\Lambda^2} \right)^{\gamma +{\rm N}} + 
                 a_0 \left( \frac{ - \Box}{\Lambda^2} \right)^{\gamma +{\rm N} -1 } \!\!\!\! +
                a_{-1} \left( \frac{ - \Box}{\Lambda^2} \right)^{\gamma +{\rm N} -2 } \!\!\!\! + \dots \right] \!
                 R^{M N}
                 \nonumber \\
                && \hspace{3.52cm}
                \equiv - 2 R_{M N}    \left[   
                \omega_{ {\rm N}} \, \Box^{\gamma +{\rm N} } + 
                 \omega_{ {\rm N} -1}   \, \Box ^{\gamma +{\rm N} -1 }  +
                 \omega_{  {\rm N} -2}   \, \Box^{\gamma +{\rm N} -2 } + \dots \right]   \! R^{M N} \, , 
\label{limiteUV}
\ee
}

where $\omega_{{\rm N}}$, $\omega_{{\rm N}-1}$ and 
$\omega_{ {\rm N}-2}$ are dimensionful constants. 
From here onwards we can apply the results of Asoreya, L$\grave{{\rm o}}$pez and Shapiro \cite{shapiro}. The one loop effective action reads,
\be
&& \Gamma^{(1 - {\rm loop})} = \frac{i}{2} {\rm Tr} \ln H_{M N, PQ} - i {\rm Tr} \ln M^{\sigma}_{\alpha} - \frac{i}{2} 
{\rm Tr} \ln C^{M N } \, , \nonumber \\
 && H_{M N, PQ}= \frac{\delta^2 S}{\delta h_{P Q} \delta h_{M N}}\Bigg|_{h=0}
 + \frac{\delta \chi_{R}}{\delta h_{P Q}} \, C^{R S } \, 
 \frac{\delta \chi_{S}}{\delta h_{M N}}\Bigg|_{h=0} 
\, .
\ee
The explicit calculation of $H$ goes behind the scope of this paper and here we only offer  
the tensorial structure in terms of the curvature tensor of the background metric and its covariant derivatives. For the action in (\ref{limiteUV}), 
the matrix $H_{M N, PQ}$ consists of three terms coming from the vertexes proportional to the 
{\em non running} constants  $\omega_{{\rm N}}$, $\omega_{{\rm N}-1}$ and $\omega_{ {\rm N}-2}$, 
\be
&& \hspace{-1.5cm} H_{M N, PQ} = H^{(\mathrm{\tilde{N} })}_{M N, PQ}+H^{(\mathrm{\tilde{N} -1})}_{M N, PQ}+H^{( \mathrm{\tilde{N} -2}) }_{M N, PQ} \,  , \nonumber \\
&& \hspace{-1.5cm} H^{(\mathrm{\tilde{N} }) }_{MN, PQ} = ( {\rm const.} \, g_{MN} g_{PQ} +{\rm permt.}) \, \Box^{\mathrm{\tilde{N}} + 2} + 
\underbrace{V^{( \mathrm{\tilde{N}}) }_{M N, PQ}\,^{L_1\dots L_{2 \tilde{\rm{N}} +2} }}_{\sim R}\nabla_{L_1} \dots \nabla_{L_{2 \tilde{\rm{N}} +2}} \nonumber \\
&& + 
\underbrace{W^{( \mathrm{\tilde{N}}) }_{M N, PQ}\,^{L_1\dots L_{2 \tilde{\rm{N}} +1} }}_{\sim \nabla R}\nabla_{L_1} \dots \nabla_{L_{2 \tilde{N} +1}} 
+ \underbrace{U^{( \mathrm{\tilde{N}}) }_{4 \, M N, PQ}\,^{L_1 \dots L_{2 \tilde{\rm{N}} } }}_{\sim R^2}\nabla_{L_1} \dots \nabla_{L_{2 \tilde{\rm{N}}}} 
\nonumber \\
&&
+ \underbrace{U^{( \mathrm{\tilde{N}}) }_{5 \, M N, PQ}\,^{L_1 \dots L_{2 \tilde{\rm{N}} -1} }}_{\sim \nabla R^2}\nabla_{L_1} \dots \nabla_{L_{2 \tilde{\rm{N}} -1}} 
+  \underbrace{U^{( \mathrm{\tilde{N}}) }_{6 \, M N, PQ}\,^{L_1\dots L_{2 \tilde{\rm{N}}-2 } }}_{\sim R^3}\nabla_{L_1} \dots \nabla_{L_{2 \tilde{\rm{N}} -2 }} \nonumber \\
&& + \dots 
+ \underbrace{U^{( \mathrm{\tilde{N}}) }_{D \, M N, PQ}\,^{L_1 \dots L_{2 \tilde{\rm{N}} +4 - D} }}_{\sim R^{D/2}}\nabla_{L_1} \dots \nabla_{L_{2 \tilde{\rm{N}} +4 - D}} \,  , 
\ee
where $\mathrm{\tilde{N}} = {\rm N} + \gamma$.
We explicitly showed the relationship of the tensors $V^{(i)}, W^{(i)}, U^{(i)}_4, U^{(i)}_5, \dots, U^{(i)}_D$
($i= \mathrm{\tilde{N}},  \mathrm{\tilde{N}-1}, \mathrm{\tilde{N}-2}$) with the curvature tensors. 
The same tensors depend also on the constants 
$\omega_{{\rm N}}$, $\omega_{{\rm N}-1}$, 
$\omega_{ {\rm N}-2}$ and $\gamma$.

Employing the universal trace formulae of Barvinsky and Vilkovisky \cite{GBV} 
\be
{\rm Tr} \ln \Box \Big|_{\rm div} \!\!\!\! & \sim & \!\! \frac{1}{\epsilon} \, \int d^D x \sqrt{|g|} \left( a_2 \, R_{MNPQ}^{\frac{D}{2}} +  a_2 \, R^{\frac{D}{2} } + \dots  \right) \, , \nonumber \\
 \nabla_{L_1} \dots \nabla_{L_p} \, \frac{1}{\Box^{ \mathrm{\tilde{N}}  + 2}} \, \delta(x,y) \Big|_{\rm div} 
\!\!\!\! &\sim&   \!\!\! \frac{1}{\epsilon} \, \left(  R_{M N P Q}^{\frac{p}{2} - (   \mathrm{\tilde{N}}  + 2) + \frac{D}{2} } 
+ \dots \right)
\,\,\,\, \,\, (p \leqslant 2 \mathrm{\tilde{N}}  + 4 )   \,  , 
\ee
we can derive 
the following divergent contribution to the effective action, 
\be 
&& 
 \Gamma^{(1)}_{\rm div}
 \sim -  \frac{1}{\epsilon}\int d^Dx \sqrt{|g|} \Big[ \beta_{\kappa} R + \beta_{\bar{\lambda}}
 + \sum_{n=0}^{\mathrm{N} } \, \Big( \beta_{a_n} R \, (-\Box_{\Lambda})^n \, R  + 
  \beta_{b_n}  \, R_{M N} \, (-\Box_{\Lambda})^n \, R^{M N}   \Big) \nonumber \\
&& \hspace{3.3cm}+
\beta_{c_1^{(1)}}  \, R^3 +  \cdots + \, \beta_{c_1^{(\mathrm{N})}} 
\, R^{\frac{D}{2}} \Big] \, ,  \label{Abeta}
\ee
where all the beta-functions $\beta_{\kappa}, \beta_{\bar{\lambda}}, \beta_{a_n}, \beta_{b_n}, \beta_{c_1^{(1)}}, \cdots, \beta_{c_1^{(\mathrm{N})}}$ depend on the constants 
$\omega_{ {\rm N}}$, $\omega_{ {\rm N}-1}$ and 
$\omega_{ {\rm N}-2}$, 
as well as on the couplings
\be
\alpha_i  \in \{\kappa, \bar{\lambda}, a_n, b_n, c_1^{(1)}, \cdots, c_1^{(\mathrm{N})} \}
\equiv 
\{\kappa, \bar{\lambda}, \tilde{\alpha}_n \} .
\ee
However, if $\gamma$ is large enough ($\gamma > D
/2$) the beta functions will depend only on 
the constants $\omega_{ {\rm N}}$, $\omega_{ {\rm N}-1}$,  
$\omega_{ {\rm N}-2}$  which specify the polynomial action in the ultraviolet regime (\ref{limiteUV}), 
but they will not depend on the coupling constants subject to renormalization. 
In other words the vertexes of the set 1 in (\ref{Vertex}) do not contribute to the beta 
functions. In this case the exponents in (\ref{alphasol}) are all equal to one ($Y_i =1$ $\forall i$) and 
$\alpha_i(\mu) \sim \alpha_i(\mu_0) + \beta_i t$. 
In particular there is no running of the cosmological constant. 

The renormalization group equations for the renormalized $\ell$-point one particle irreducible 
vertices of the dimensionless field $g_{MN}$ give: 
\be
\Gamma^{(\ell)} (\mu \, q,  \kappa^{-2}, \tilde{\alpha}_n, \mu_0 ) = \mu^{2 \gamma + 2 \mathrm{N}+4} \, 
\Gamma^{(\ell)} (q,  \mu^{-(2 \gamma + 2 \mathrm{N}+2)} \,  \kappa^{-2}, 
\mu^{-(2 \gamma + 2 \mathrm{N})+ 2n} \, \tilde{\alpha}_n(\mu), \mu_0 ) 
\ee
where $\mu_0$ is the renormalization mass and $q$ is the Euclidean momentum. 
Ignoring the inessential tensor structure, for any $\ell$ these vertices read 
\be
\Gamma^{(\ell)} \sim \kappa^{-2} \, q^2 +  
\sum_n^{\mathrm{N}} a_n \, q^{2 n +4} + \sum_n^{\mathrm{N}} b_n \, q^{2 n +4} 
+ c \, q^{2 {\rm N} + 4} 
\, \ln \left( \frac{q^2}{\mu_0^2} \right)  
+ \dots + \omega_{ \mathrm{N} } \,
q^{2 \gamma + 2 \mathrm{N}+4} \, , 
\label{gamman}
\ee
where the last term comes from the large $z$ limit in the entire functions $h_i(z)$ and ``$c$" is a constant. 
Therefore, under $q \rightarrow \mu \, q$, (\ref{gamman}) gives: 
\be
 && \hspace{-0.7cm}
 \Gamma^{(\ell)} \! \sim \! \mu^{2 \gamma + 2 \mathrm{N}+4}  
\Big[ \mu^{-(2 \gamma + 2 \mathrm{N}+2)} \, \kappa^{-2} \, q^2 +  
\sum_{n=0}^{\mathrm{N}}   \mu^{-(2 \gamma + 2 \mathrm{N} ) + 2n }  \,   a_n(\mu) \, q^{2 n +4} 
+ \sum_{n=0}^{\mathrm{N}}    \mu^{-(2 \gamma + 2 \mathrm{N}) + 2n  }  \,   b_n(\mu) \, q^{2 n +4} 
\nonumber \\
&& \hspace{-0.0cm}
+  c \, \mu^{- 2 \gamma} 
 \, q^{2 \mathrm{N}+4} \, \ln \left( \frac{q^2}{\mu_0^2} \right)  
+ \dots + 
\omega_{ \mathrm{N} } \, q^{2 \gamma + 2 \mathrm{N}+4} 
 +\mu^{-2} \,  \omega_{ \mathrm{N} -1} \, q^{2 \gamma + 2 \mathrm{N}+2}
  + \mu^{-4} \, \omega_{ \mathrm{N} -2 } \, q^{2 \gamma + 2 \mathrm{N}} 
\Big] .
\label{gammant}
\ee
We see that for $\mu \rightarrow + \infty$ the $q^{2 \gamma + 2 \mathrm{N}+4}$ terms 
completely dominate, and all the other couplings including the Newton constant $\kappa$ become 
irrelevant as expected in an asymptotically free theory \cite{TomboAF}. 
Finally, let us consider the effective action in the ultraviolet limit. 
The following expansion around the flat spacetime,  
\be
g_{MN} = \eta_{MN} +  \left( \frac{\Lambda}{ \mu}   \right)^{\gamma +  {\rm N}+2 } \!\!\! 
h_{MN}
 : = \eta_{MN} + f \, h_{MN} \, ,  \nonumber \\
\ee
greatly simplifies the action (\ref{action}) in the ultraviolet regime ($\mu \rightarrow + \infty$), namely 
%
\be
\mathcal{L} \simeq   
h \, \Box^{\gamma +{\rm N} +2} \, h + f \, h \, h \, \Box^{\gamma + {\rm N} +2} h + \dots =
h \, \Box^{\gamma +{\rm N} +2}\, h + O(f^2) \,  .
\ee
The letter limit 
implies that the main terms of the action are the kinetic parts of the non-polynomial 
operators $R h_0(\Box_{\Lambda})R$ and $R_{MN} h_2(\Box_{\Lambda}) R^{MN}$
and that the asymptotic freedom in $f$ leads to the validity of the perturbation theory in $f$ itself.

\section{Appendix C. More about the form factor}\label{AC}
The form factor used in this paper and suggested in \cite{Tombo} can be written in the following 
form which makes the polynomial asymptotic behavior explicit,
\be
&& \hspace{-1.22cm}
V^{-1}(z) =  e^{\frac{1}{2} 
\Gamma \left(0, p_{\gamma + \mathrm{N} + 1}^2(z) \right)  } \,  
e^{\frac{\gamma_E}{2}} \,
\left| p_{\gamma + \mathrm{N} + 1}(z) \right|  
\nonumber \\
&&
=  \underbrace{
e^{\frac{\gamma_E}{2}} \,
\left| p_{\gamma + \mathrm{N} + 1}(z) \right| 
}_{V^{-1}_{\infty}(z) } +
    \underbrace{ 
\left(   e^{\frac{1}{2}  \Gamma \left(0, p_{\gamma + \mathrm{N} + 1}^2(z) \right)  }  -1 \right) 
e^{\frac{\gamma_E}{2}} 
\, \left| p_{\gamma + \mathrm{N} + 1}(z) \right|  
}_{V^{-1}(z) -V^{-1}_{\infty}(z) \, \sim \, e^{- F(z)}  \,\,, \,\, \,\, F(z)>0} \, .
\label{Vlimit3}
\ee
The first correction to such limit 
goes to zero faster than any polynomial function, 
namely 
\be
\lim_{z \rightarrow +\infty} V(z)^{-1} = e^{\frac{\gamma_E}{2}} \, |z|^{\gamma + \mathrm{N} +1} \,\,\,\, 
\,\,\,\, {\rm and} \,\,\,\, \,\,\,
\lim_{z \rightarrow +\infty} 
\left(\frac{V(z)^{-1}}{e^{\frac{\gamma_E}{2}} |z|^{\gamma + \mathrm{N} +1} } - 1 \right) z^n = 0
\,\,\,\, \forall \, n \in \mathbb{N}\, .
\label{propertyAppendix}
\ee
%
For $p_{\gamma + \mathrm{N} + 1}(z) =z^{\gamma + \mathrm{N} + 1} \equiv z^n$ 
($n \equiv \gamma + \mathrm{N} + 1$),
the function $F(z)$ in (\ref{Vlimit3})
is well approximated, at least along the real axis, by $F(z) \propto z^m$ with 
$m \in \mathbb{N}$, $m \gtrsim n$.
Two examples are: 
$F(z) \approx 2 |z|^5$ for $n=4$ or 
$F(z) \approx 2 z^{12}$ for $n=10$. 

For 
N$=0$ and $\gamma =3$ (in $D=4$), 
\be
V^{-1}(z)  \sim  e^{\frac{e^{-z^8}}{2 z^8}}\, e^{\frac{\gamma _E}{2}}    \, z^4 
=    e^{\frac{\gamma _E}{2}}    \, z^4 + \underbrace{ \left( e^{\frac{e^{-z^8}}{2 z^8}} -1 \right) \, e^{\frac{\gamma _E}{2}}    \, z^4}_{e^{-F(z)}}
 \,\,  ,
\ee
whereas for $p_{\gamma + \mathrm{N} + 1}(z) =z^{n}$ 
($n \equiv \gamma + \mathrm{N} + 1$) 
the asymptotic behavior is:
\be
V^{-1}(z)  \sim  
e^{\frac{e^{-z^{2 n} } }{2 z^{2 n} } \left(\dots \,  -120 z^{-10 n}+24 z^{-8 n}-6 z^{-6 n}+2 z^{-4
   n}-z^{-2 n}+1\right)}    \, e^{\frac{\gamma _E}{2}}
(-\Box)^n    \, .
\ee
We can rewrite the form factor by separating the leading order from the next one, 
namely 
\be
&& \hspace{-1.6cm} 
V^{-1}(- \Box) \sim  
e^{\frac{e^{-\Box^{2 n} } }{2 \Box^{2 n} } \, \left(\dots \,  -\frac{120}{ \Box^{10 n}}+
\frac{24}{\Box^{8 n}} - \frac{6}{\Box^{6 n}} + \frac{2}{\Box^{4 n}} - \frac{1}{\Box^{2 n}} +1\right)} \, 
e^{\frac{\gamma _E}{2}}  \, (-\Box)^n 
\nonumber \\
&& = \underbrace{e^{\frac{\gamma _E}{2}}  \,
(-\Box)^n}_{V^{-1}_{\infty}(-\Box)  } + 
\underbrace{  \left(e^{\frac{e^{-\Box^{2 n} } }{2 \Box^{2 n} } \, \left(\dots \,  -\frac{120}{ \Box^{10 n}}+
\frac{24}{\Box^{8 n}} - \frac{6}{\Box^{6 n}} + \frac{2}{\Box^{4 n}} - \frac{1}{\Box^{2 n}} +1\right)} -1  \right)
e^{\frac{\gamma _E}{2}}  \,
(-\Box)^n 
}_{V^{-1}_{{\rm next} - \infty}(-\Box)  } \, , 
\ee 
where $\Box$ necessarily means $\Box/\Lambda^2$. 
For any fixed number of gravitons, 
the polynomial leading contribution $V^{-1}_{\infty}(-\Box)$ gives a finite number of vertex operators,
which are responsible for the one-loop divergences. 

Let us concentrate on the order that follows the leading one, $V^{-1}_{{\rm next} - \infty}(-\Box)$. 
In order to 
expand the form factor in powers of the graviton fluctuation, 
we extract the d'Alembertian operator around the flat spacetime from the general covariant one,
\be
&&\hspace{-2.17cm} V^{-1}_{{\rm next} - \infty}(-\Box) =  \left( e^{\frac{e^{-\big(\Box_{\eta} - (\Box_{\eta} + \Box) \big)^{2 n} } }{2 \big(\Box_{\eta} - (\Box_{\eta} + \Box) \big)^{2 n} } + \dots} -1 \right)
e^{\frac{\gamma _E}{2}}  \,
\Big( -\Box_{\eta}  + ( \Box_{\eta} - \Box) \Big)^n 
 \nonumber \\
%
&& = 
 \left( e^{\frac{e^{-\Box_{\eta}^{2 n} } }{2 \Box_{\eta}^{2 n} } } -1 \right) 
  ( 1 + \dots) \, e^{\frac{\gamma _E}{2}}  
  \left( (-\Box_{\eta})^n+ O\left( (\partial^2 h)^n \right) + \dots  \right)
\nonumber \\
&& = 
\left( e^{\frac{e^{-\Box_{\eta}^{2 n} } }{2 \Box_{\eta}^{2 n} } } -1 \right)
 \, e^{\frac{\gamma _E}{2}}
 \left( (-\Box_{\eta})^n+ O\left( (\partial^2 h)^n \right) + \dots  
\right) \, .
\label{asy3}
\ee
In (\ref{asy3}) $\Box_{\eta} = \eta_{\mu \nu} \partial^{\mu} \partial^{\nu}$ and the ``$\dots$" on the left in the second line refer to graviton vertexes coming from the Taylor expansion of the exponential, 
being careful not to overlook 
the commutators  $[ \Box, \Box_{\eta}] R_{\dots} \neq 0$. 
Due to the presence of the fast convergent exponential pre-factor in (\ref{asy3}), we do not 
get any divergent contribution to the effective action coming from $V^{-1}_{{\rm next} - \infty}(-\Box)$.
However, the infinite number of vertexes with a fixed number of external gravitons, will give 
a non trivial contribution to the finite part of the effective action. This task is beyond the scope of this paper.



%

\section{Acknowledgments}
A special thanks goes to Damiano Anselmi and Ilya Shapiro for the numerous 
insights into quantum field theory. 
We also thank Jorge Russo, Yuri Gusev, Sergei Odintsov, Emilio Elizalde.

\end{document}